\documentclass[preprint, superscriptaddress]{revtex4-1}
\usepackage[dvips]{graphicx}

\begin{document}

\title{Realization of a reversible switching in TaO$_{2}$ polymorphs via Peierls distortion for resistance random access memory}
\date{\today}
\author{Linggang Zhu}
\affiliation{School of Materials Science and Engineering, Beihang University, Beijing 100191, China}
\affiliation{Center for Integrated Computational Materials Engineering, International Research Institute for Multidisciplinary Science, Beihang University, Beijing 100191, China}
\author{Jian Zhou}
\affiliation{School of Materials Science and Engineering, Beihang University, Beijing 100191, China}
\author{Zhonglu Guo}
\affiliation{Center for Integrated Computational Materials Engineering, International Research Institute for Multidisciplinary Science, Beihang University, Beijing 100191, China}
\affiliation{College of Materials, Xiamen University, Xiamen 361005, China}
\author{Zhimei Sun}
\email[Corresponding author: ]{zmsun@buaa.edu.cn}
\affiliation{School of Materials Science and Engineering, Beihang University, Beijing 100191, China}
\affiliation{Center for Integrated Computational Materials Engineering, International Research Institute for Multidisciplinary Science, Beihang University, Beijing 100191, China}

\begin{abstract}

Transition-metal-oxide based resistance random access memory is a promising candidate for next-generation universal non-volatile memories. Searching and designing appropriate new materials used in the memories becomes an urgent task. Here, a new structure with the TaO$_{2}$ formula was predicted using evolutionary algorithms in combination with first-principles calculations. This new structure having a triclinic symmetry (\emph{T}-TaO$_{2}$) is both energetically and dynamically more favorable than the commonly believed rutile structure (\emph{R}-TaO$_{2}$). Our hybrid functional calculations show that \emph{T}-TaO$_{2}$ is a semiconductor with a band gap of 1.0 eV, while \emph{R}-TaO$_{2}$ is a metallic conductor. This large difference in electrical property makes TaO$_{2}$ a potential candidate for resistance random access memory (RRAM). Furthermore, we have shown that \emph{T}-TaO$_{2}$ is actually a Peierls distorted \emph{R}-TaO$_{2}$ phase and the transition between these two structures is via a directional displacement of Ta atoms. The energy barrier for the reversible phase transition from \emph{R}-TaO$_{2}$ to \emph{T}-TaO$_{2}$ is 0.19 eV/atom and the other way around is 0.23 eV/atom, suggesting low power consumption for the resistance switch. The present findings provide a new mechanism for the resistance switch and will also stimulate experimental work to fabricate tantalum oxides based RRAM.

\end{abstract}
\maketitle

Faster and denser memories are ever-increasing demands and requirements of the ongoing advances in electronic devices and multimedia applications, etc~\cite{Han2013}. One of the most promising candidates for next generation non-volatile memories is resistance random access memories (RRAM), where data storage is achieved by the switch between a high resistance state (logic 0) and a low resistance state (logic 1) in certain materials by applying electric pulses. The advantages of RRAM over the current commercial silicon-based Flash memory and other memory types are its simple structure, good scalability, fast switching speed and good compatibility with the widely used complementary metal-oxide-semiconductor (CMOS) technologies~\cite{Hiro2012}. In a RRAM device, the choice of recording materials is important to achieve the above performances, and hence, searching and designing novel materials suitable for RRAM are received considerable focus. Numerous materials have been demonstrated to have the resistance switch behavior, among which the most promising type is transition-metal-oxides (TMO), such as HfO$_{x}$~\cite{Shang2014}, TiO$_{x}$~\cite{Hu2014}, TaO$_{x}$~\cite{Lee2011, Park2013, Kim2014} and so on. The typical RRAM device involving TMO has a sandwich-like Metal/TMO/Metal structure, where the TMO layer is not restricted to one single type of oxide, for instance, the Pt/SiO$_{2}$/Ta$_{2}$O$_{5-x}$/TaO$_{2-x}$/Pt device has been recently developed by Samsung Electronics~\cite{Park2013}. However, the microscopic mechanism of resistance switch in TMO-RRAM devices is very complex and not easy to unravel. Previously, various mechanisms have been proposed~\cite{Han2013, Pan2014}, including filamentary conduction, space charge and traps, charge transfer, ionic conduction, etc. The filamentary conduction mechanism induced by the formation/migration of point defects has been reported in many TMO-RRAM~\cite{Gu2010, Tohru2011}. For example, the diffusion of oxygen vacancies was found responsible for the resistance switch behavior in Ta$_{2}$O$_{5-x}$/TaO$_{2-x}$, where the resistance of the intrinsically insulating Ta$_{2}$O$_{5}$ can be tuned by the generation of oxygen vacancies and clustering of Ta ions and while the metallic layer TaO$_{2-x}$ mainly serves as the oxygen vacancy/ion reservoir~\cite{Park2013}. Nevertheless, the complex microscopic mechanism of resistance switch in RRAM limits its practical applications. For large scale applications, obviously, a simple and controllable microscopic resistance switch in recording materials is vital. This is also the dream of materials scientists and a hot topic in RRAM research field.

The above mentioned TaO$_{2}$ was treated as a metallic material. However, in this work, we report a new stable triclinic structured TaO$_{2}$ (\emph{T}-TaO$_{2}$) semiconductor with a band gap as large as 1.0 eV, in contrast to the metallic nature of the TaO$_{2}$ with rutile structure that people normally believed. This new phase can be reversibly switched with its polymorph conducting phase (\emph{R}-TaO$_{2}$) via Peierls distortion where the phase transition energy barrier is 0.19 eV/atom from \emph{R}-TaO$_{2}$ to \emph{T}-TaO$_{2}$ and is 0.23 eV/atom for the other way around. This huge resistance difference and easy phase transition between the two TaO$_{2}$ polymorphs make TaO$_{2}$ a promising candidate for RRAM. Furthermore, the present results provide a new resistance switching mechanism for RRAM, i.e., the metal-insulator phase transition via Peierls distortion, thus this work also expands the catalog of RRAM materials. The present findings open a new way to design novel TaO$_{2}$-based RRAM devices with lower power consumption, where the microscopic resistance switching is simple and controllable.

\textbf{Results}

\begin{figure}
\protect{\includegraphics[width=0.9\columnwidth]{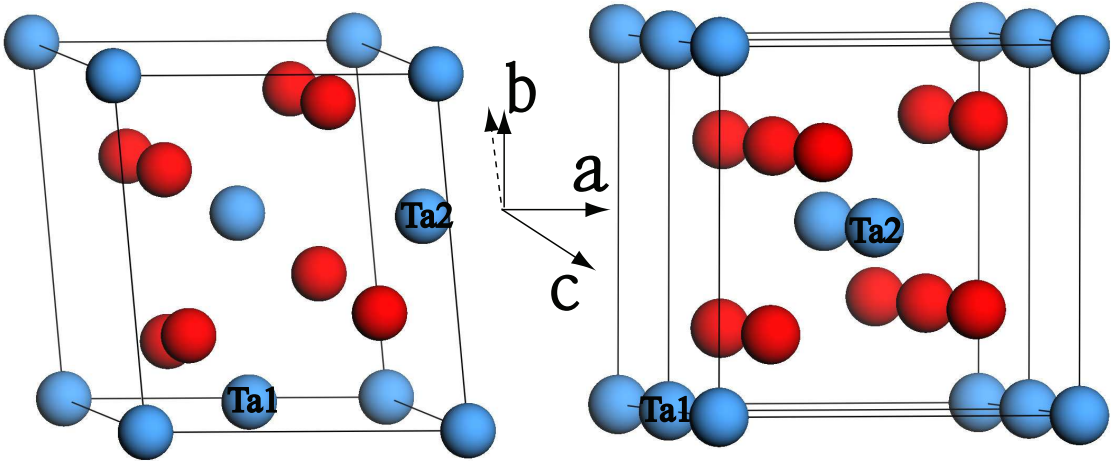}}
\caption{(color online) Left: Unit cell of \emph{T}-TaO$_{2}$; Right: $1\times1\times2$ supercell of \emph{R}-TaO$_{2}$. Blue and red balls represent Ta and O, respectively. Ta atoms showing the most significant displacements in the two structures are labeled as Ta1 and Ta2.}
\label{fig:AtomStruc}
\end{figure}

\textbf{Geometry and properties of the newly found \emph{T}-TaO$_{2}$.} Since some substoichiometric oxides exist or might be formed in the resistance-switch process of RRAM, the structure and property of any crystalline or even amorphous compounds formed by the transition metal and oxygen are vital for investigating the switch mechanism and designing novel RRAM devices. Here, using tantalum oxide as an example, we make a thorough exploration of the structures of oxide Ta(IV)O$_{2}$. TaO$_{2}$ is generally believed to have a rutile (tetragonal) structure (\emph{R}-TaO$_{2}$)~\cite{Nils1954, Syono1983, Garg1996}, which is more stable than another symmetry \emph{I4$_{1}$/a}~\cite{Romanov2009}. The present evolutionary structure search and ab initio calculations (computational settings can be found in the Method section) revealed a new stable triclinic TaO$_{2}$ phase (symmetry group \emph{P-1}), for which the lattice parameters are: $\alpha$ = 90.0$^\circ$, $\beta$ = 87.3$^\circ$, $\gamma$ = 95.7$^\circ$; a = 4.70~\AA, b = 5.20~\AA, c = 5.72~\AA. Details of the atomic structures can be found in the supplementary material. This newly found geometry contains 12 atoms in one unit cell, i.e., 4 TaO$_{2}$ formula units, as shown in Fig.~\ref{fig:AtomStruc}. To make a clear comparison, a $1\times1\times2$ supercell consisting of 12-atoms for \emph{R}-TaO$_{2}$ (symmetry group \emph{P42/mnm}) was also shown. The lattice parameters for this \emph{R}-TaO$_{2}$ supercell are: $\alpha$ = $\beta$ = $\gamma$ = 90$^\circ$; a = b = 4.99~\AA, c = 5.76~\AA. It is obvious that the most distinct difference between these two structures is the periodicity in the c direction, where for \emph{T}-TaO$_{2}$ it is almost twice that of \emph{R}-TaO$_{2}$. While the deviation in lattice parameters along the a and b directions is within 6\%. Further close analysis of the atomic positions in the two structures shown in Fig.~\ref{fig:AtomStruc} reveals that, besides the distortion of the octahedron formed by six O atoms, the labeled Ta atoms have the most significant displacements and these two inequivalent atoms displace almost half length of lattice parameter a. This large displacement results in a doubled periodicity along the c direction for \emph{T}-TaO$_{2}$. Therefore, \emph{T}-TaO$_{2}$ can be regarded as a distorted structure from the high-symmetric \emph{R}-TaO$_{2}$. The phase transition from \emph{R}-TaO$_{2}$ to \emph{T}-TaO$_{2}$ is induced by a typical Peierls distortion.

\begin{figure}
\centering
\includegraphics[width=0.8\columnwidth]{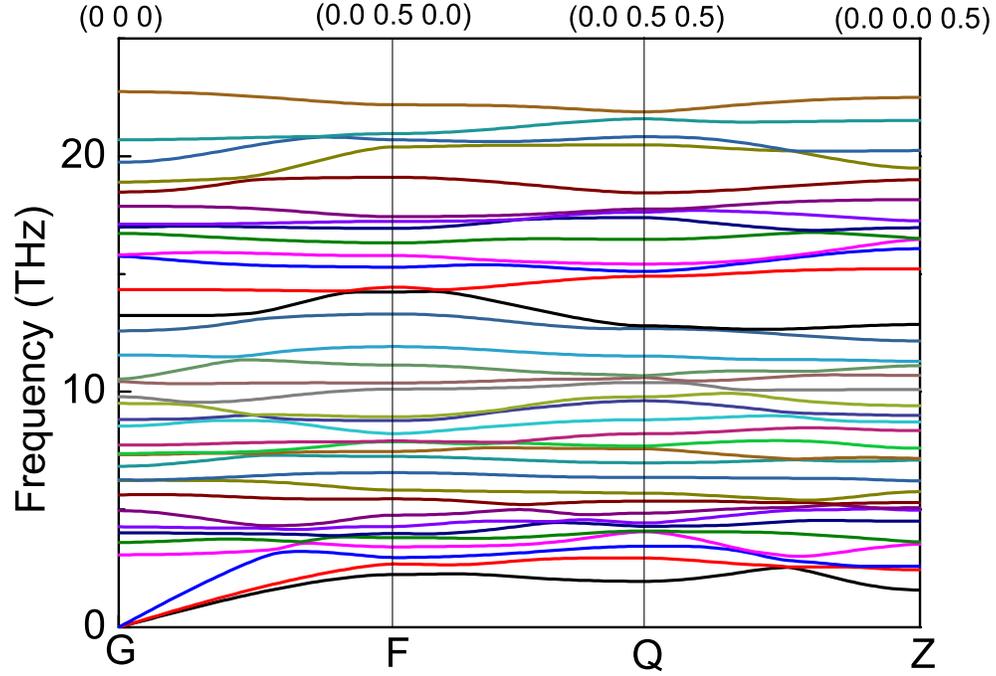}\\
(a)\\
\includegraphics[width=0.8\columnwidth]{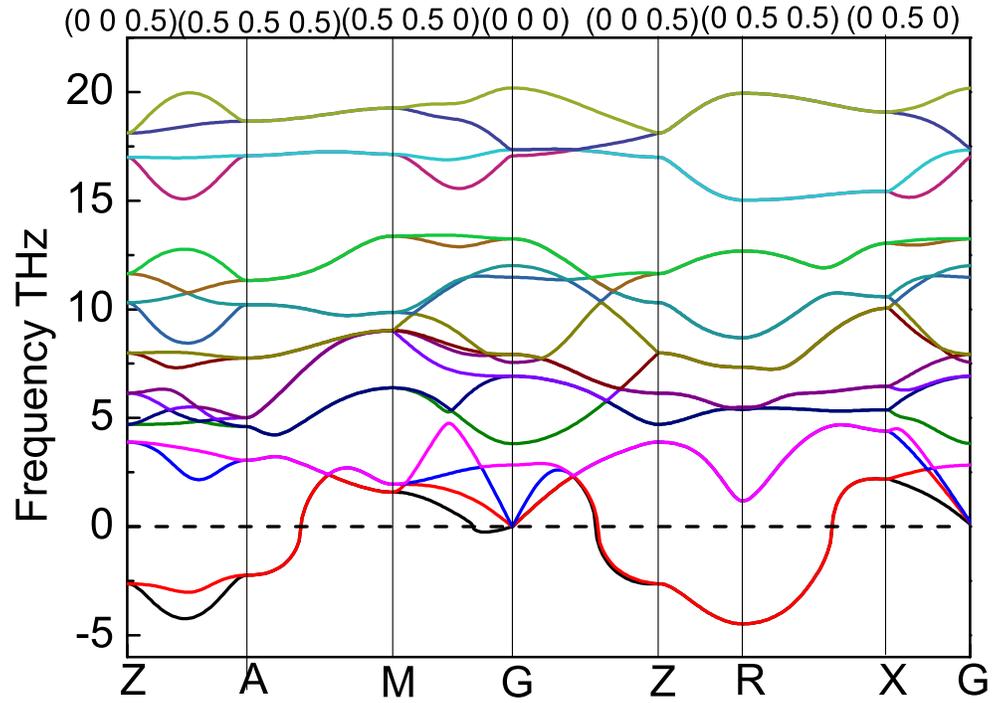} \\
(b)\\
\caption{\label{fig:Phonon} Phonon dispersion for (a) \emph{T}-TaO$_{2}$ and (b)\emph{R}-TaO$_{2}$, the black dashed line indicating the position of zero frequency.}
\end{figure}

The phase stability of the two TaO$_{2}$ polymorphs are studied by the calculated cohesive energy and their phonon dispersions. Based on our PAW-GGA calculations, the cohesive energy of \emph{T}-TaO$_{2}$ is 0.12 eV per formula unit lower than \emph{R}-TaO$_{2}$. While our hybrid functional calculations showed that the cohesive energy of \emph{T}-TaO$_{2}$ is 0.20 eV per formula unit lower than \emph{R}-TaO$_{2}$. This indicates that the exchange effect of electron interactions is quite different for these two phases with the same composition. Nevertheless, \emph{T}-TaO$_{2}$ is energetically more favorable than \emph{R}-TaO$_{2}$. The results also clearly show that the structure distortions of \emph{R}-TaO$_{2}$ lead to a more stable geometry. The phonon dispersions shown in Fig.~\ref{fig:Phonon} clearly unravel that \emph{T}-TaO$_{2}$ is also dynamically stable, while \emph{R}-TaO$_{2}$ is not as shown by the imaginary frequencies spreading over a large range of reciprocal space in Fig.~\ref{fig:Phonon}(b). Therefore, the present found \emph{T}-TaO$_{2}$ phase is both energetically and dynamically more stable than \emph{R}-TaO$_{2}$.

\begin{figure}
\centering
\includegraphics[width=0.8\columnwidth]{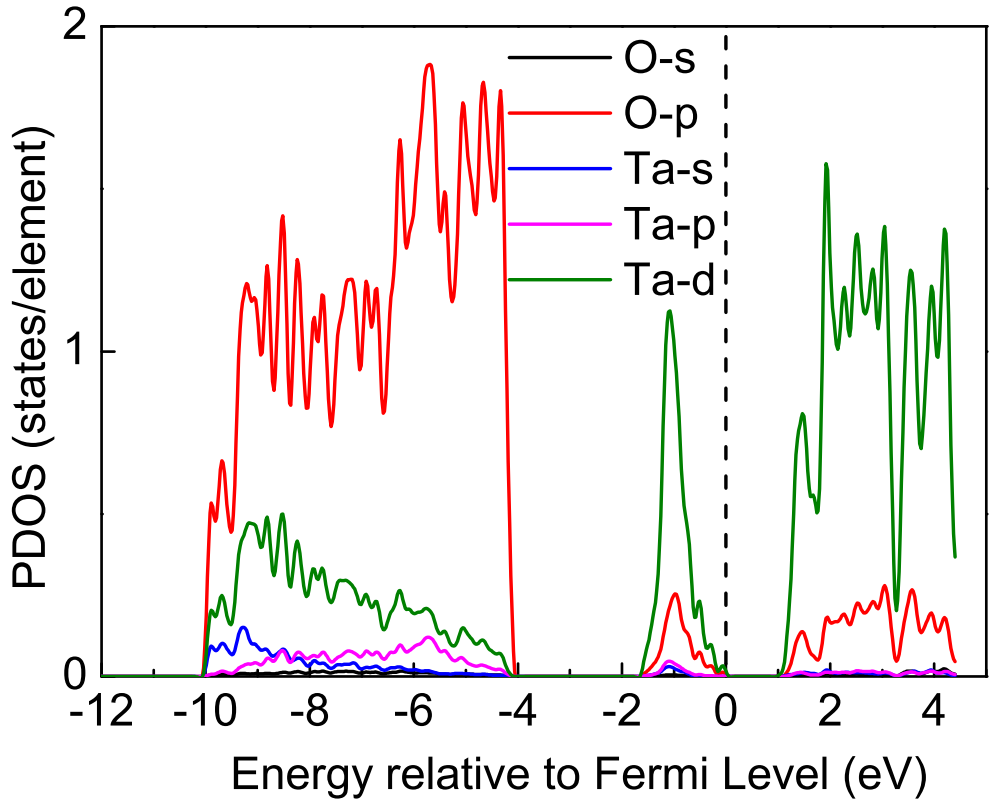}\\
(a)\\
\includegraphics[width=0.8\columnwidth]{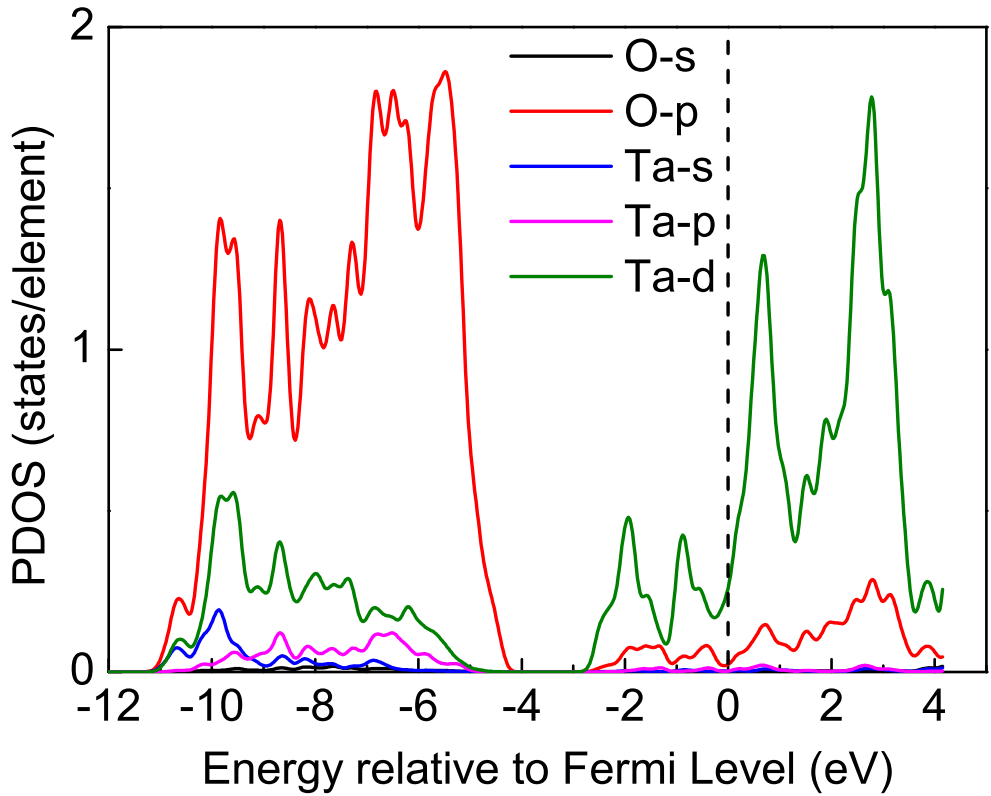} \\
(b)\\
\caption{\label{fig:PDOS} Partial density of states for (a) \emph{T}-TaO$_{2}$ and (b)\emph{R}-TaO$_{2}$. The vertical dashed lines indicate the position of Fermi level.}
\end{figure}

To investigate the electronic property of the two TaO$_{2}$ polymorphys, the partial density of states (PDOS) were calculated using hybrid functional, the results of which are shown in Fig.~\ref{fig:PDOS}. It is clear that \emph{T}-TaO$_{2}$ is a semiconductor with a band gap as large as 1.0 eV, while \emph{R}-TaO$_{2}$ is metallic with no band gap. Furthermore, for both structures, the top of the valence-band mainly consists of the overlapping states of Ta-\emph{d} and O-\emph{p}, where the difference is whether these overlapped states are localized (for \emph{T}-TaO$_{2}$) or delocalized (for \emph{R}-TaO$_{2}$). This significant difference in the electronic property of these two polymorphs makes TaO$_{2}$ a potential candidate material for RRAM devices, where \emph{T}-TaO$_{2}$ is a high resistance state (HRS) and \emph{R}-TaO$_{2}$ is a low resistance state (LRS). As mentioned above, \emph{T}-TaO$_{2}$ can be regarded as a distorted \emph{R}-TaO$_{2}$. This distortion leads to a more stable and doubled periodicity in the c direction and hence results in the metal-insulator transition (MIT), which we believe is a typical Peierls transition, similar to our previous work in GeTe system~\cite{Sun2012}.

\begin{figure}
\protect{\includegraphics[width=1.0\columnwidth]{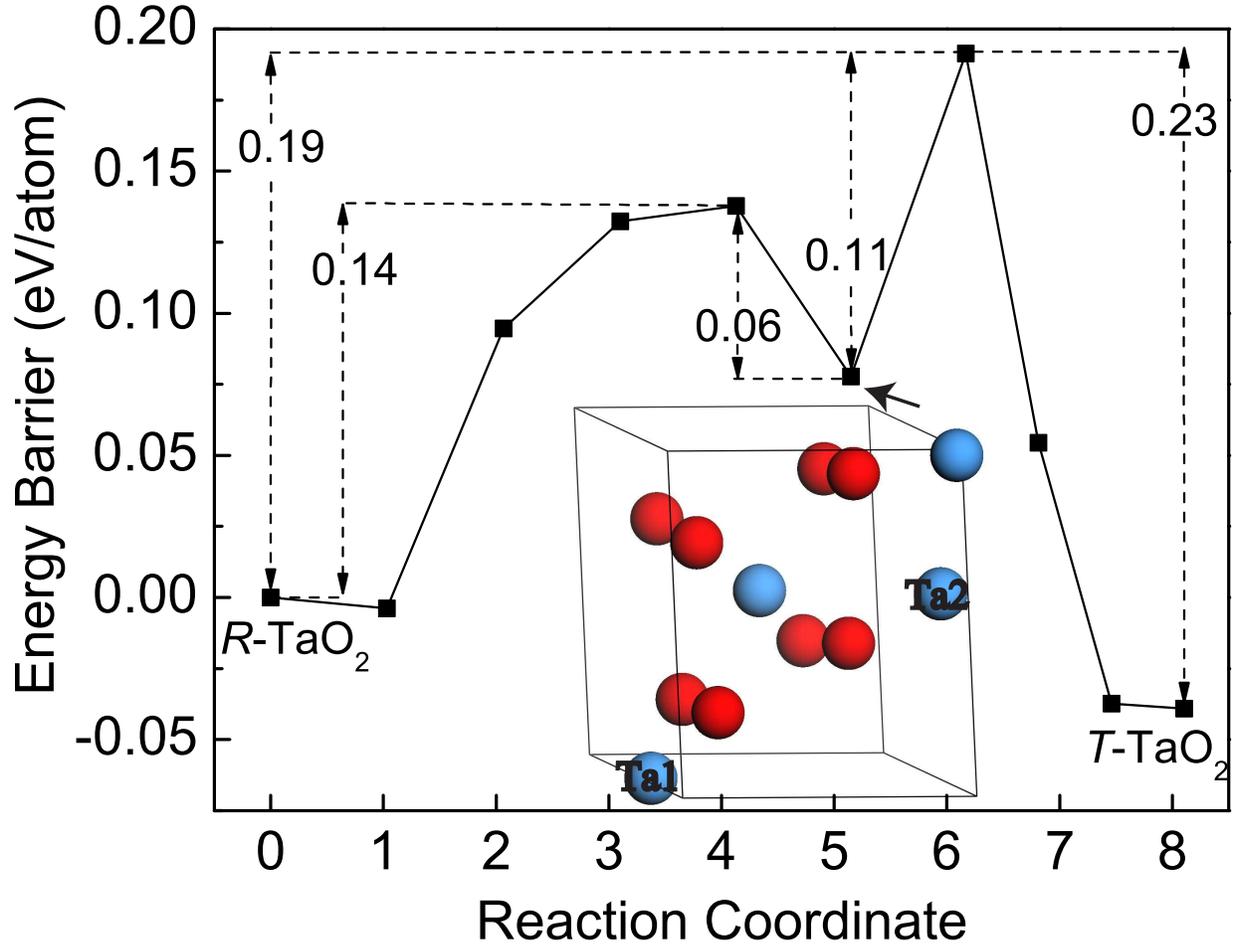}}
\caption{Transition path from \emph{R}-TaO$_{2}$ to \emph{T}-TaO$_{2}$, the energy difference between various states is labeled in the unit of eV/atom. Structure of the meta-stable intermediate state as denoted by the solid arrow is also shown, where the Ta atoms corresponding to numbered ones in Fig.~\ref{fig:AtomStruc} are labeled as Ta1 and Ta2.}
\label{fig:NEB}
\end{figure}

\textbf{Phase transition between \emph{R}-TaO$_{2}$ and \emph{T}-TaO$_{2}$ and the induced switching process.} For an efficient RRAM device, as required by the read/write speed and power consumption, switch between high and low resistance states should be fast and easy. To explore the possibility of \emph{T}-TaO$_{2}$ to \emph{R}-TaO$_{2}$ transition being the new candidate for a RRAM device, we have calculated the phase transition process, and the phase transition path and energy barrier are illustrated in Fig.~\ref{fig:NEB}. Note that the first intermediate state next to \emph{R}-TaO$_{2}$ was found to be slightly more stable than \emph{R}-TaO$_{2}$, showing distortion induced stability which is consistent with its dynamically unstable nature. Interestingly, as shown in Fig.~\ref{fig:NEB}, the transition path between \emph{R}-TaO$_{2}$ and \emph{T}-TaO$_{2}$ has two distinct periods, separated by a meta-stable state with its structure also shown in the figure. Compared the metastable phase with the two end structures shown in Fig.~\ref{fig:AtomStruc}, especially the positions of the two numbered Ta atoms,  we arrive at the conclusion that these two Ta atoms play the pivotal role in the MIT where they move independently but directionally with the largest migration distance to accomplish the phase transition between \emph{R}-TaO$_{2}$ and \emph{T}-TaO$_{2}$. During the transition from \emph{R}-TaO$_{2}$ (LRS) to \emph{T}-TaO$_{2}$ (HRS), atom Ta2 firstly migrates a large distance to the similar symmetry position as in \emph{T}-TaO$_{2}$ leading  to a meta-stable state (the inset of Fig.~\ref{fig:NEB}), where the energy barrier to reach this state is 0.14 eV/atom. Then it is followed by a large migration of atom Ta1 leading to the structure of \emph{T}-TaO$_{2}$, where the energy barrier for this meta-stable phase transforming to \emph{T}-TaO$_{2}$ is 0.11 eV/atom. The overall barrier to overcome for the MIT transition from \emph{R}-TaO$_{2}$ to \emph{T}-TaO$_{2}$ is around 0.19 eV/atom. The other way around is for the phase transition from \emph{T}-TaO$_{2}$ (HRS) to \emph{R}-TaO$_{2}$ (LRS) as applying an opposite electric field, where the overall energy barrier is 0.23 eV/atom. At this stage, it is obvious that TaO$_{2}$ is a suitable candidate material for RRAM devices, where the reversible switch between HRS and LRS is a Peierls distortion type which can be achieved by applying different electric fields of opposite directions and various energies. We believe this new TaO$_{2}$-RRAM should be low power consumption. Even though there is no direct relation between the energy barrier of this metal-semiconductor transition with the power consumption of real devices, it is reasonable that the larger the energy barrier, the higher the power consumption. The conclusion is obvious as compared TaO$_{2}$ with other materials also used in random access memory. For example, in the well-known Ge-Sb-Te (GST) system for phase-change random access memory (PCRAM) where the reversible phase transition between amorphous (HRS) and crystalline (LRS) under various electric pulse is used for data storage~\cite{Lencer2008, Sun2006, Sun2007, Sun2009}, Kolobov et al.~\cite{Kolobov2011} recently reported that the energy barrier between the ordered-crystalline and the distorted structure leading to the amorphous state is as large as 0.45 eV/atom, almost twice that of the barrier in our study. Therefore, we believe that the power consumption when using the metal-semiconductor transition between the two TaO$_{2}$ polymorphs in RRAM could be lower than that of GST in PCRAM. Moreover, like the filamentary mechanism that commonly exists in the RRAM based on TMO, the resistance switch between \emph{R}-TaO$_{2}$ and \emph{T}-TaO$_{2}$ normally occurs in a tiny area of the material in practical RRAM devices, not necessarily inside the whole bulk. This will result in an even lower energy consumption and faster data-storage speed, i.e., make the device 'greener'. Finally, it is worth to mention that the above calculated negative phonon frequencies in \emph{R}-TaO$_{2}$ will not affect the application of this material in RRAM. A similar case is in Ge$_{2}$Sb$_{2}$Te$_{5}$, a prototype material for PCRAM, where the resistance switch is accomplished by the reversible transition between a metastable rocksalt structure and an amorphous state. For metastable rocksalt structured Ge$_{2}$Sb$_{2}$Te$_{5}$ negative phonon frequencies have also been reported by theoretical calculations~\cite{Tsafack2011}.

\textbf{Conclusion}

In summary, by using evolutionary algorithms, we found a new tantalum dioxide (\emph{T}-TaO$_{2}$) structure, which is more stable than the commonly believed rutile symmetry (\emph{R}-TaO$_{2}$). The newly found \emph{T}-TaO$_{2}$ which can be regarded as distorted \emph{R}-TaO$_{2}$ is a semiconductor, in contrast to the metallic nature of \emph{R}-TaO$_{2}$. This Peierls distortion induced reversible metal-semiconductor transition with low energy barriers of 0.19 and 0.23 eV/atom, respectively, provides a new mechanism for RRAM devices and expands the catalog of RRAM materials. Finally, our work will stimulate the re-building of transition-metal-oxide phase diagram aiming at finding new structures with various electronic properties, and also promote the experimental work to fabricate RRAM devices based on these transition metal oxides.

\textbf{Methods}

\textbf{Structure prediction.} Evolutionary algorithm as implemented in the code USPEX~\cite{Oganov2006, Oganov2011} was used to predict the structures with the formula TaO$_{2}$. During the calculations, the already known rutile TaO$_{2}$ structure was introduced as the seed.

\textbf{Ab initio calculations.} All the structures screened by USPEX were fully relaxed using VASP code~\cite{Kresse1993, Kresse1996, KresseG1996}, in conjunction with projector augmented wave potentials within the generalized-gradient approximation (PAW-GGA)~\cite{Perdew1996}, while for the calculation of the band gap of structures, hybrid functional HSE06~\cite{Heyd2006} was employed to deal with the localized \emph{d} electrons of Ta. Pseudopotentials with electronic configurations of $5p^65d^36s^2$ and $2s^22p^4$ were used for Ta and O, respectively. The structures mentioned in this paper were fully relaxed until the forces on them were less than 10 meV/\AA~with a cutoff energy 450 eV. \emph{K}-point meshes for structures with different symmetry and lattice size were generated with the same resolution. Our calculations also showed that the spin-polarization effects in the structures involved in the present study were negligible.
Phonon dispersions were calculated by Density Functional Perturbation Theory (DFPT) as implemented in the code Phonopy~\cite{Togo2008}. The phase transition process was investigated using the generalized solid-state Nudged Elastic Band (G-SSNEB) method~\cite{Sheppard2012}. First, we assumed a linear transition path by sampling 8 intermediate states between the two end-structures (\emph{R}-TaO$_{2}$ and \emph{T}-TaO$_{2}$), and during the Minimum Energy Path (MEP) optimization, besides the atomic positions, the shape of the supercells is also allowed to relax.

\textbf{Acknowledgement}

This work is partially supported by National Natural Science Foundation for Distinguished Young Scientists of China (51225205) and the National Natural Science Foundation of China (61274005, 37691801).

\textbf{Author contributions}

Z.S. and J.Z. supervised the study. L.Z and Z.G performed the calculations. L.Z, J.Z, and Z.S. analyzed the data and wrote the paper.

\textbf{Additional information}

\textbf{Competing financial interests:} The authors declare no competing financial interests.

\end{document}